\input harvmac.tex
\noblackbox
%


\def\unlockat{\catcode`\@=11}
\def\lockat{\catcode`\@=12}

\unlockat

\def\newsec#1{\global\advance\secno by1\message{(\the\secno. #1)}
\global\subsecno=0\global\subsubsecno=0\eqnres@t\noindent
{\bf\the\secno. #1}
\writetoca{{\secsym} {#1}}\par\nobreak\medskip\nobreak}
\global\newcount\subsecno \global\subsecno=0
\def\subsec#1{\global\advance\subsecno
by1\message{(\secsym\the\subsecno. #1)}
\ifnum\lastpenalty>9000\else\bigbreak\fi\global\subsubsecno=0
\noindent{\it\secsym\the\subsecno. #1}
\writetoca{\string\quad {\secsym\the\subsecno.} {#1}}
\par\nobreak\medskip\nobreak}
\global\newcount\subsubsecno \global\subsubsecno=0
\def\subsubsec#1{\global\advance\subsubsecno by1
\message{(\secsym\the\subsecno.\the\subsubsecno. #1)}
\ifnum\lastpenalty>9000\else\bigbreak\fi
\noindent\quad{\secsym\the\subsecno.\the\subsubsecno.}{#1}
\writetoca{\string\qquad{\secsym\the\subsecno.\the\subsubsecno.}{#1}}
\par\nobreak\medskip\nobreak}

\def\subsubseclab#1{\DefWarn#1\xdef
#1{\noexpand\hyperref{}{subsubsection}%
{\secsym\the\subsecno.\the\subsubsecno}%
{\secsym\the\subsecno.\the\subsubsecno}}%
\writedef{#1\leftbracket#1}\wrlabeL{#1=#1}}
\lockat

%
%
%

\def\CK{{\cal K}}

\def\CM{{\cal M}}
\def\CN{{\cal N}}

\def\Det{{\rm Det}}

\def\IZ{\relax\ifmmode\mathchoice
{\hbox{\cmss Z\kern-.4em Z}}{\hbox{\cmss Z\kern-.4em Z}}
{\lower.9pt\hbox{\cmsss Z\kern-.4em Z}}
{\lower1.2pt\hbox{\cmsss Z\kern-.4em Z}}\else{\cmss Z\kern-.4em
Z}\fi}
\def\IB{\relax{\rm I\kern-.18em B}}
\def\IC{{\relax\hbox{$\inbar\kern-.3em{\rm C}$}}}
\def\ID{\relax{\rm I\kern-.18em D}}
\def\IE{\relax{\rm I\kern-.18em E}}
\def\IF{\relax{\rm I\kern-.18em F}}
\def\IG{\relax\hbox{$\inbar\kern-.3em{\rm G}$}}
\def\IGa{\relax\hbox{${\rm I}\kern-.18em\Gamma$}}
\def\IH{\relax{\rm I\kern-.18em H}}
\def\II{\relax{\rm I\kern-.18em I}}
\def\IK{\relax{\rm I\kern-.18em K}}
\def\IP{\relax{\rm I\kern-.18em P}}

\def\inbar{\,\vrule height1.5ex width.4pt depth0pt}
\def\p{\partial}
\def\pb{{\bar \p}}

\font\cmss=cmss10 \font\cmsss=cmss10 at 7pt
\def\IR{\relax{\rm I\kern-.18em R}}

\def\Tr{\rm Tr}

\def\hone{h^{(1)} }
\def\lgdt#1#2{\log\Det' \Delta^{#1, #2} }

%
\lref\fhsv{
S. Ferrara, J. A. Harvey, A. Strominger, C. Vafa ,
``Second-Quantized Mirror Symmetry, '' Phys. Lett. {\bf B361} (1995) 59;
hep-th/9505162. }

\lref\hm{J. A. Harvey and G. Moore,
``Algebras, BPS states, and strings,''
Nucl. Phys. {\bf B463}(1996)315; hep-th/9510182.}

\lref\dkl{L. Dixon,  V. S. Kaplunovsky and J. Louis, ``Moduli-dependence of
string
loop corrections to gauge coupling constants, ''Nucl. Phys. {\bf B329} (1990)
27. }

\lref\kaplouis{V. Kaplunovsky and J. Louis, ``On gauge couplings in string
theory,''
Nucl. Phys. {\bf B444} (1995) 191, hep-th/9502077. }

\lref\agn{I. Antoniadis,  E. Gava, K.S. Narain, ``Moduli corrections to
gravitational
couplings from string loops,'' Phys. Lett. {\bf B283} (1992) 209,
hep-th/9203071; `` Moduli corrections to gauge and gravitational couplings in
four-dimensional superstrings,'' Nucl. Phys. {\bf B383} (1992) 109,
hep-th/9204030.}

\lref\agnt{I. Antoniadis,
 E. Gava, K.S. Narain and T.R. Taylor,
``Superstring threshold corrections to
Yukawa couplings,'' Nucl. Phys {\bf B407} (1993) 706;
hep-th/9212045. Note: These versions are
different. }

\lref\monstref{B. H. Lian and S. T. Yau, ``Arithmetic properties of mirror map
and quantum
coupling, '' hep-th/9411234. }

\lref\lianyau{B. H. Lian and S. T. Yau, ``Arithmetic properties of mirror map
and quantum
coupling, '' hep-th/9411234;
``Mirror Maps, Modular Relations
and Hypergeometric Series I ,'' hep-th/9506210;
``Mirror Maps, Modular
Relations and Hypergeometric Series II,''
hep-th/9507153}

\lref\cfetc{S. Cecotti, S. Ferrara, L. Girardello, A. Pasquinucci , M. Porrati
, ``Matter coupled supergravity with
Gauss-Bonnet invariants: Component Lagrangian
and supersymmetry breaking,'' Int. J. Mod. Phys. {\bf A3} (1988)
1675.}
\lref\carlustov{G.L. Cardoso, D. Lust, and
B.A. Ovrut, ``Moduli dependent non-holomorphic
contributions of massive states to gravitational
couplings and $C^2$ terms in $Z(N)$ orbifold compactification,''
Nucl. Phys. {\bf B436} (1995) 65; hep-th/9410056.}

\lref\rey{ G. L. Cardoso, G. Curio,  D. Lust,  T. Mohaupt, S.-J. Rey,
``BPS Spectra and Non--Perturbative Couplings in
N=2,4 Supersymmetric String Theories,'' Nucl. Phys. {\bf B464} (1996) 18;
hep-th/9512129.}
\lref\kiretal{E. Kiritsis, C. Kounnas, P.M. Petropoulos,
J. Rizos, hep-th/9605011}
\lref\tkawai{Toshiya Kawai , ``String Duality and Modular Forms,''
hep-th/9607078.}

\lref\henning{M. Henningson  and   G. Moore,
``Threshold Corrections in $K3\times T2$ Heterotic String Compactifications,''
hep-th/9608145.}
\lref\lustline{G. L. Cardoso,  G. Curio,  D. Lust,
``Perturbative Couplings and Modular Forms in N=2 String Models with a Wilson
Line,''
hep-th/9608154.}

\lref\bcov{M. Bershadsky, S. Cecotti, H. Ooguri and C. Vafa, `` Kodaira-Spencer
theory
of gravity and exact results for quantum string amplitudes, '' Commun. Math.
Phys.
{\bf 165} (1994) 311, hep-th/9309140. }

\lref\klt{V. Kaplunovsky, J. Louis and S. Theisen, ``Aspects of duality in
$N=2$ string vacua,'' Phys. Lett. {\bf B357} (1995) 71, hep-th/9506110.}

\lref\louispas{J. Louis, PASCOS proceedings, P. Nath ed., World
Scientific 1991.}

\lref\lco{G. L. Cardoso and B. A. Ovrut, ``A Green-Schwarz mechanism
for $D=4$, $N=1$ supergravity anomalies,'' Nucl. Phys. {\bf B369} (1992) 351;
``Coordinate and Kahler sigma model anomalies and their cancellation
in string effective field theories,''  Nucl. Phys. {\bf B392} (1993) 315,
hep-th/9205009.}

\lref\borchiv{R. E. Borcherds, ``The moduli space
of Enriques surfaces and the fake monster Lie
superalgebra,'' Topology vol. 35 no. 3, (1996) 699.}

\lref\borchmoon{R. E. Borcherds, ``Monstrous moonshine and monstrous Lie
superalgebras,'' Invent. Math. {\bf 109} (1992) 405.}

\lref\moonshine{J. H. Conway and S. P. Norton, ``Monstrous moonshine,''
Bull. London Math. Soc. {\bf 11} (1979) 308. }

\lref\borchvii{R. Borcherds, ``Automorphic forms
with singularities on Grassmannians,''
alg-geom/9609022.}

\lref\nikulin{see e.g. V. V. Nikulin, ``Reflection groups in hyperbolic spaces
and the denominator formula for Lorentzian Kac-Moody algebras,
alg-geom/9503003.}

\lref\flm{I. B. Frenkel, J. Lepowsky, and A. Meurman, {\it Vertex operator
algebras
and the monster,} Pure and Applied Mathematics Volume 134, Academic
Press, San Diego, 1988.}

\lref\asplouis{P. Aspinwall and J. Louis,
``On the Ubiquity of K3 fibrations in
in string duality,'' Phys. Lett. {\bf B369} (1996) 233; hep-th/9510234.}

\lref\klm{A. Klemm, W. Lerche and P. Mayr, ``K3-fibrations and
Heterotic-Type II string duality, '' Phys. Lett. {\bf B357}
(1995) 313, hep-th/9506122.}

\lref\cogp{P. Candelas, X. de la Ossa, P. S. Green and L. Parkes,
``A pair of Calabi-Yau manifolds as an exactly soluble
superconformal theory,'' Nucl. Phys. {\bf B359} (1991) 21.}

\lref\effcomp{  C .  Vafa, ``Evidence for F-Theory,'' hep-th/9602022;
 D. R. Morrison,  C. Vafa, ``Compactifications of F-Theory on
Calabi--Yau Threefolds -- I,'' hep-th/9602114;
D. R. Morrison,  C. Vafa, ``Compactifications of F-Theory on
Calabi--Yau Threefolds -- II,'' hep-th/9603161.}

\lref\clash{A. Losev, G. Moore, N. Nekrasov,
and S. Shatashvili,
``Chiral Lagrangians, Anomalies, Supersymmetry, and Holomorphy,''
hep-th/9606082.}

\lref\bcova{M. Bershadsky, S. Cecotti, H. Ooguri and C. Vafa,
``Holomorphic anomalies in topolgical field theories,'' Nucl. Phys.
{\bf B405} (1993) 279; hep-th/9302103.}

\lref\hmiii{J. A. Harvey and G. Moore, ``Five-brane instantons and
$R^2$ couplings in $N=4$ string theory,'' hep-th/9610237.}

\lref\hmalg{J. A. Harvey and G. Moore, ``On the algebra of BPS
states,'' hep-th/9609017.}

\lref\aspintran{P. Aspinwall, ``An $N=2$ dual pair and a phase transition,''
Nucl. Phys. {\bf B460} (1996) 57;
hep-th/9510142.}

\lref\jtii{J. Jorgenson and A. Todorov, ``Enriques surfaces, analytic
discriminants, and Borcherd's $\Phi$ function,'' Yale preprint.
}

\lref\vafatest{C. Vafa, ``A stringy test of the fate of the conifold,''
Nucl. Phys. {\bf B447} (1995) 252; hep-th/9505023.}

\lref\swb{N. Seiberg and E. Witten, `` Monopoles, duality and chiral symmetry
breaking in $N=2$ supersymmetric QCD, '' Nucl. Phys. {\bf B431} (1994) 484,
hep-th/9408099. }

%
%

\Title{\vbox{\baselineskip12pt
\hbox{hep-th/9611176}
\hbox{EFI-96-40 }
\hbox{YCTP-P22-96 }
}}
{\vbox{\centerline{Exact Gravitational Threshold
Correction}
\centerline{in the FHSV model
 } }}

\centerline{Jeffrey A. Harvey}
\bigskip
\centerline{\sl Enrico Fermi Institute, University of Chicago}
\centerline{\sl 5640 Ellis Avenue, Chicago, IL 60637 }
\centerline{\it harvey@poincare.uchicago.edu}
\bigskip
\centerline{Gregory Moore}
\bigskip
\centerline{\sl Department of Physics, Yale University}
\centerline{\sl New Haven, CT  06511}
\centerline{ \it moore@castalia.physics.yale.edu }

\bigskip
\centerline{\bf Abstract}

We consider the automorphic forms
which govern the gravitational threshold correction $F_1$
in models of heterotic/IIA
duality with $N=2$ supersymmetry in four dimensions.
In particular we derive the
full nonperturbative formula for $F_1$ for the dual pair
originally considered by Ferrara, Harvey, Strominger and
Vafa (FHSV).   The answer involves
an interesting automorphic product
constructed by Borcherds which is associated to
the ``fake Monster Lie superalgebra.''
As an application of this result we rederive
a result of Jorgenson \& Todorov on
determinants of $\bar \partial$ operators on
$K3$ surfaces.

\Date{Nov. 18, 1996}
%

\newsec{Introduction}

In previous work we have  developed the idea that
there is an algebraic structure which underlies many dualities
in string theory \refs{\hm, \hmalg}. This point of view is
supported by the fact that threshold corrections in $N=2$
theories are given by sums of logarithmic functions
over the roots of generalized
Kac Moody algebras (GKM) \hm\ and by the fact that one can define an algebra
on BPS states which in some cases is closely related to a
 generalized Kac-Moody algebra
\hmalg. Since threshold corrections are determined purely by the
spectrum of BPS states, there must be a relation between these two
facts. Unfortunately the precise relation is
still not clear. In fact, to our knowledge the denominator product of
a known GKM has not yet appeared in a calculation of threshold corrections
in any model
(such products have appeared as {\it part} of the
answer in several papers).
 In this paper we will remedy this by showing that the
denominator function of the
``fake Monster Lie superalgebra'' \borchmoon\ governs
the gravitational threshold correction $F_1$ in the $N=2$ dual pair
of FHSV \fhsv.

In the second and third sections of this paper we discuss some general
features of  gravitational threshold corrections in $N=2$ heterotic
and type IIA dual pairs. In the fourth section we apply these results and
second quantized mirror symmetry to determine the exact non-perturbative
form for $F_1$ in the FHSV dual pair. The fifth section applies this result
to rederive a result of Jorgenson and Todorov on the determinant of
$\bar \partial$ operators on $K3$ surfaces which double cover the
Enriques surface. We make a few brief comments on the relation to
F theory in section 6 and conclude in section 7.

\newsec{Gravitational threshold correction
for $d=4,\CN=2$ heterotic compactifications}

\subsec{One-loop integral}

We first consider
$d=4,N=2$ heterotic compactifications
with the ten-dimensional gauge group broken to some rank $s$ subgroup
(there are a total of $s+4$ vector gauge fields).
The tree level vectormultiplet moduli space is
therefore:
\eqn\vmmod{
\eqalign{
\CM_{vm} & = \{ (\tau_S,y) \} \in SL(2,\IR)/SO(2) \times \CN^{s+2,2} \cr
\CN^{s+2,2}& = O(\Gamma^{s+2,2})\backslash O(s+2,2)/[O(s+2)\times O(2)] . \cr
 }}
String and automorphic fields $S, \tau_S$,
respectively are defined as follows. We choose $Re(S)>0$ and
define
\eqn\quess{
q_S \equiv e^{ - 8 \pi^2 S } = e^{ 2 \pi i \tau_S}
}
which is invariant under axion shifts. We normalize the Yang-Mills
fields so that the action is
${1 \over  2 g^2} \int \tr F \wedge *F $ with
$S={1 \over  g^2} + i { \theta \over  8 \pi^2} $.

In this paper we focus on the effective, non-Wilsonian
coupling
\eqn\gbcoup{
\int_{\IR^{1,3}} {1 \over  24 g^2_{\rm grav}(p^2)}
\tr R\wedge R^* .
}
At string tree level $ g^{-2}_{\rm grav}(p^2) = 24 Re(S)$.
These couplings have been widely discussed in
the literature
\refs{\agn, \cfetc, \carlustov, \bcov, \klt, \rey, \kiretal}.
A one-loop formula for the derivatives of $g_{\rm grav}$ with
respect to the vector moduli was derived in \agn:
\eqn\dercoup{
{\p \over  \p y^i} \biggl( {1 \over  g^2_{\rm grav}(p^2)}\biggr)
= {\p \over  \p y^i}{\Delta^{(1)}_{\rm grav} \over  16 \pi^2}
}
where in the above we are holding the string coupling $S$ fixed.
The quantity $\Delta^{(1)}_{\rm grav}$ is given in \agn\ as
\eqn\gc{
\Delta^{(1)}_{\rm grav}  = \int_{\cal F} {d^2 \tau \over \tau_2}
\left[ {- i   \over \eta^2 (\tau)} {\rm Tr}_R
\left\{J_0 e^{i \pi J_0} q^{L_0 - c / 24}
\bar{q}^{\bar{L}_0 - \bar{c} / 24}
\left(E_2 (\tau) -
 {3 \over \pi \tau_2}\right) \right\} -
b_{\rm grav} \right]
}
where the trace is over the Ramond sector, $J_0$ is the $U(1)$ current
of the $N=2$ superconformal algebra, and $E_2$ is the second Eisenstein
series with a $q$ expansion
\eqn\etwoexp{E_2 = 1 - 24 \sum_{n=1}^\infty \sigma(n) q^n}
with $\sigma(n)$ the sum of the divisors of $n$. The constant
$b_{\rm grav}$ is the ``gravitational beta function'' of \agn\
and ensures that the integral \gc\ is finite away from enhanced
symmetry subvarieties.
{}From \gc\ we may obtain the duality invariant
effective coupling:
\eqn\agntgc{
\bigl({1 \over  g^2_{\rm grav}(p^2)}\bigr)^{HET}=
 {24 \over  g^2_{\rm string} }
+
{b_{\rm grav} \over  16 \pi^2} \log {M_{\rm Planck}^2 \over  p^2}
+
{\Delta^{(1)}_{\rm grav} \over  16 \pi^2}
- {5 n_V + n_H \over  8 \pi^2} \log [ Re S]
}
where $ g^{-2}_{\rm string}  = Re(S) + \Delta^{\rm univ}/ 16 \pi^2$, $n_H$ is
the number of massless hypermultiplets,
and $n_V$ is the number of massless vectormultiplets (including the
graviphoton). The quantity $\Delta^{\rm univ}$ is related to a Green-Schwarz
term for sigma-model anomalies.
We  have chosen to write the second term in \agntgc\ with the
Planck mass rather than the string scale in the
logarithm since it is the Planck mass that is duality invariant.
Actually, \agntgc\ does not quite follow from \dercoup. The last term in
\agntgc,  which affects only the dilaton
one-point function, is very difficult to compute and is not determined
by \dercoup.
\foot{Deriving the last term in \agntgc\
from a low energy field theory analysis
is an extremely challenging problem.
We thank V. Kaplunovsky and J. Louis
for detailed discussions about this.}
It may however be determined by appealing to heterotic-IIA string duality.
In particular it can be determined from the expressions we derive in
section 3 for the corresponding quantities in  IIA theory.

Now we derive two important properties of
the integral \gc.  From \hm\ we have the
formula:
\eqn\vmhm{
\eqalign{
{i\over 2} {1 \over  \eta^2} \Tr_R J_0 e^{i \pi J_0}
q^{L_0 - c/24}
&
\bar q^{\tilde L_0 - \tilde c/24}
= \cr
  \Biggl[
\sum_{\rm BPS\  vectormultiplets}
q^{\Delta} \bar q^{\bar \Delta}
-&
\sum_{\rm BPS\  hypermultiplets}
q^{\Delta} \bar q^{\bar \Delta} \Biggr] \cr}
}
This implies:

\item{1.}
Using the expansion  \etwoexp\ and noting that \vmhm\ starts with $q^{-1} +
\cdots$
we see that  the constant is
\eqn\gvrbet{
b_{\rm grav} = 2( n_H - n_V + 24)
}
thus recovering the result of \agn.

\item{2.} If we approach an enhanced symmetry
variety at which masses of BPS vector multiplets, $m_v$, and/or
masses of BPS hypermultiplets, $m_h$ approach zero
then the integral \gc\ picks up a divergence
\eqn\massdiv{
2 \biggl[
\sum_{\rm BPS\  vectormultiplets} \log(4 \pi m_v^2) -
\sum_{\rm BPS\  hypermultiplets} \log(4 \pi m_h^2)
\biggr]
}

\subsec{Evaluating the integral}

Using the techniques of
\refs{\dkl,\hm} the above integral can be
evaluated for ``rational backgrounds''
which satisfy:
\eqn\ratback{
{i \over  2} {1 \over  \eta^2} \Tr_R J_0 e^{i \pi J_0}
q^{H}
\bar q^{\tilde H}
= \sum_i Z^i_\Gamma(q,\bar q) f_i(q)
}
where $f_i(q)$ form a unitary representation
of $SL(2,\IZ)$ and $Z_i$ are a set of partition
functions for a lattice and its translates.

The evaluation of \gc\ was done in
\refs{\hm, \tkawai, \henning, \lustline, }  with
the following result.
Introduce the pseudo-invariant
dilaton:
\eqn\thrshiv{
\eqalign{
\tilde S  & = S +  {1 \over  s+4  }  \eta^{ab}{\p \over  \p y^a}
{\p \over  \p y^b} \hone \cr}
}
such that $Re(\tilde S)$ is invariant under
$T$-duality. In the above $\hone$ is the one-loop prepotential.
Then we have at 1-loop:
\foot{The strange prefactor is meant to define a
nice modular form. It also facilitates the comparison
with the IIA string.}
\eqn\dfnfrm{
\eqalign{
{4 \pi^2 \over  3} \bigl({1 \over  g^2_{\rm grav}(p^2)}\bigr)^{HET}
&=
{b_{\rm grav} \over  12}
\log {M_{\rm Planck}^2 \over  p^2} -
\log\parallel \Psi_{grav}^{\leq 1} (S,y) \parallel^2 \cr}
}
where
\eqn\invlog{
\eqalign{
\log\parallel \Psi_{grav}^{\leq 1} (S,y) \parallel^2 & =
- 4 Re\bigl[ 8 \pi^2 \tilde S \bigr] + {5 n_V + n_H \over  6}
\log[Re S]
+ \log \parallel \Psi_{\rm grav}(y) \parallel^2\cr
\log \parallel \Psi_{\rm grav}(y) \parallel^2
& = \log \vert \Psi_{\rm grav}(y) \vert^2
 + { b_{grav} \over  12} (\log[- (Im y)^2] - \CK)\cr}
}
and
\eqn\startform{
 \Psi_{grav}^{\leq 1}(S,y)
= q_{\tilde S}^2 \Psi_{grav}(y).
}
The quantity $\CK$ in \invlog\ is an unimportant constant defined in \hm.
Each term on the RHS of \dfnfrm\ is $T$-duality
invariant.
$T$-duality invariance and \massdiv\ lead
to the following key properties of
$\Psi_{\rm grav}(y)$:

\item{1.} $\Psi_{\rm grav}(y)$ is a modular form of
$O(s+2,2;\IZ)$ of weight
\eqn\autweight{
w= {b_{\rm grav}\over  12} =4 - {n_V - n_H \over  6} .
}

\item{2.} If $y$ approaches a subvariety
where there are new massless vector and hyper multiplets
then the LHS of \dfnfrm\ diverges like:
\eqn\massdivp{
{1 \over  6} \biggl[
\sum_{vm} \log(4 \pi m^2) - \sum_{hm} \log(4 \pi m^2)
\biggr]
}
Therefore, if $\tilde S$ does not pick up a divergence,
and if there are
$\Delta n_V$ vector multiplets and  $\Delta n_H$ hypermultiplets
with masses vanishing to first order in the distance to the subvariety
then $\Psi_{\rm grav}(y)$ obtains a zero of order:
\eqn\ordzero{
{ 1 \over  6} [ \Delta n_H - \Delta n_V]
}

\subsec{Loop corrections}

Finally, let us discuss higher loops.
The above result \dfnfrm\ can be
cast in the form:
\eqn\holomorph{
\eqalign{
\bigl({1 \over  g^2_{\rm grav}(p^2)}\bigr)^{HET}
&=
Re\bigl[  \CF_1^{HET}(S,y) \bigr] +
{b_{\rm grav} \over  16 \pi^2}
\biggl[ \log {M_{\rm Planck}^2 \over  p^2}
+ K^{(0)}(S, \bar S, y, \bar y) \biggr]\cr
&  - {3 \over  4 \pi^2} (4-n_V)
\log[Re S]\cr}
}
where $K^{(0)}$ is the tree level Kahler potential,
$K^{(0)} = -log Re S - log[-(Im y)^2]$.
$\CF_1^{HET}(S,y)$ is holomorphic, and has
transformation properties such that
$ g^{-2}_{\rm grav}(p^2)$ is invariant.
As discussed in \klt,
$N=2$ supersymmetry says that
\eqn\ntooc{
  \CF_1^{HET}(S,y) = 24 S +
  \CF_1^{HET,(1)} (y) + \CO(e^{- 8 \pi^2 S})
}
so \dfnfrm\ implies the formula for $\CF_1^{HET}(S,y)$
to all orders of perturbation
theory. Moreover, as the notation suggests,
we expect that $\Psi_{grav}^{\leq 1} (S,y)$
in \startform\
should be regarded as the first term of an
expansion in $q_S$ of an
automorphic product on the
exact vectormultiplet moduli space. Thus, it should be
possible to understand the $\log[Re S]$
correction purely within the heterotic
string by requiring $T$-duality invariance
of the invariant norm-square
$\parallel \Psi(S,y)\parallel^2$
of the full product. However, this might
require a nonperturbative understanding of
certain infrared effects.

\newsec{Gravitational threshold correction
for $d=4,\CN=2$ type II compactification}

We now turn to gravitational threshold corrections in type II
string theory on a Calabi-Yau 3-fold $X_3$ with Kahler potential
$K$, Betti numbers $h^{1,1}$ and $h^{2,1}$ and Euler number
$\chi = 2(h^{1,1} - h^{2,1})$.

The Gauss-Bonnet coupling for  the type II
string on a Calabi-Yau 3-fold $X_3$
is given by \refs{\bcova, \bcov}:
\eqn\agntgc{
{4 \pi^2 \over  3}
\bigl({1 \over  g^2_{\rm grav}(p^2)}\bigr)^{II}=
 {b_{\rm grav} \over  12} \log {M_{\rm Planck}^2 \over  p^2}
+
F_1^{kahler}
}
where the last term is obtained
from a   fundamental domain
integral:
\eqn\bcoviii{
\eqalign{
F_1 & = \int_{\CF} {d^2 \tau \over \tau_2}
\Biggl[
\Tr_{RR} (-1)^{J_L} J_L  (-1)^{J_R} J_R
q^H \bar q^{\tilde H} - const\Biggr]\cr
& = F_1^{cplx} + F_1^{kahler}\cr}
}
Here the trace is over the Ramond-Ramond sector and $J_L$, $J_R$
are the left and right-moving $U(1)$ currents of the $(2,2)$ superconformal
algebra.
As shown in \bcov, $F_1$
splits as a sum that depend only on complex and
Kahler moduli respectively. These terms    are
exchanged by mirror symmetry.

The formula analogous to
\holomorph\
was given in \bcov:
 \eqn\bcovf{
\eqalign{
F_1^{II} & = \log\Biggl[e^{(3+h_{1,1}-\chi/12)K}
(\det K_{i \bar j})^{-1} \vert {1 \over  \Psi_1^{II }(t)}\vert^2  \Biggr]\cr
& = (3 + h_{1,1} - {\chi \over  12} ) K - \log \det K_{i\bar j}
- 2 \log \vert  \Psi_1^{II }(t)\vert \cr
& \equiv - \log \parallel \Psi_{grav}^{II}(t)\parallel^2  \cr}
}

The type II  result analogous to \massdiv\
was obtained in \vafatest. Massless particles
lead to a singularity in $F_1$ of:
\eqn\massii{
{1 \over  6} \bigl[\sum_{vm} \log m^2 - \sum_{hm} \log m^2\bigr]
}
In order to go further we must specialize to Calabi-Yau spaces
$X_3$ which have heterotic duals. This requires that $X_3$ be
a K3 fibration \refs{\klm, \asplouis}.

\subsec{Application to K3 fibrations}

We now assume $X_3$ is a $K3$ fibration
and follow the discussion of Aspinwall
and Louis \asplouis.
We choose integral generators $e_s, e_a$ of
$H^2(X_3;Z)$ and introduce coordinates $(t_s,t_a)$
on the complexified
Kahler form:
\eqn\comkah{
\omega= t_s e_s + \sum_{a=1}^{s+2} t_a e_a
}
where $e_s$ is dual to the K3 fiber.
The $e_a$ form a basis for integral
$(1,1)$ classes for the K3 fiber.
The Picard lattice $Pic(K3)$ will have signature $(1,s+1)$
and  the  $t_\alpha$ are coordinates for the
Kahler cone in $Pic(K3)\otimes \IC$.
Therefore, we may write:
\eqn\counting{
\eqalign{
n_V & = s+4\cr
3 + h_{1,1} - \chi/12 & = (12+5 n_V + n_H)/6\cr}
}

The coordinates $t_s, t_a$
are flat (special) coordinates and
in \asplouis\ it is shown that we have
the identification:
\eqn\idcoords{
\eqalign{
t_s & = 4 \pi i S   \cr
t_a & = y^a  \cr}
}
The holomorphic
anomaly at large $Im(t_s) $ is computed from
\eqn\corrkay{
\eqalign{
K & = K^{(0)} + \CO(1/Im(t_s) ) \cr
K^{(0)} & = - \log[Im(t_s)] - \log[-(Im t_\alpha)^2] + const. \cr}
}
giving:
\eqn\bcovfii{
\eqalign{
F_1^{II} & = - {  5 n_V + n_H \over  6} \log[Re S ]
+ {4 \pi^2 \over  3} 24 Re(S) \cr
& + {n_V - n_H - 24 \over  6} \log[-2 (Im t_\alpha)^2] \cr
& - \log\vert\Psi^{II}(t_\alpha) \vert^2 + \CO(e^{2 \pi i t_s})\cr}
}
The fact that there are only
exponential corrections is surprising since
we expect
 $K,\log \det K_{i \bar j} $ to be corrected to all orders in
$1/Im(t_s)$. However, it is implied by  \ntooc.

We conclude that at large $Re(S)$ we have:
\eqn\tpeiipsi{
\Psi_{grav}^{II}(t_s, t_a) = q_{\tilde S}^2 \Psi_{grav}^{II}(t_a) (1+
\CO(q_S) )
}

Now, $F_1^{II}$ must
be an invariant function on the moduli
space. Combined with \massii\
we obtain two conditions which
are the analogs of the heterotic
conditions \autweight\ and \ordzero:

\item{1.} $\Psi_{grav}^{II}(t_\alpha)$ is a modular form of
weight
\eqn\weight{
{1 \over 12} b_{grav}=4-\chi/12
}

\item{2.}  The zeroes of $\Psi^{II}_{\rm grav}(t_a)$
at an enhanced symmetry
point with hypermultiplet and vectormultiplet masses
vanishing to first order are
of order:
\eqn\zeros{
 {1 \over  6} [\Delta n_H  -  \Delta n_V ]
}
when  $\tilde S$ is well-defined.

Now we can apply the Koecher principle:
Modular forms on $\CN^{s+2,2}$ are completely
determined by their weight, zeroes and poles.
In general we need to allow for phases (i.e.,
a ``multiplier system'') in
the modular transformation laws. However,
such phases correspond to one-dimensional
representations of the T-duality group
$O(\Gamma)$. It follows from a theorem
of Kazhdan
\ref\kazhdan{C. Delaroche and A. Kirillov,
``Sur les relations entre l'espace dual d'un
groupe et la structure de ses sous-groupes
ferm\'es,'' Sem. Bourbaki, 1968, p. 343.}\  that
the abelianization $O(\Gamma)/[O(\Gamma),O(\Gamma)]$
is always a finite group, and therefore,
we can eliminate the phases by raising
the form to an appropriate power.
\foot{For the example considered in the
next section the power $4$ is sufficient. }
 Comparing \weight\zeros\ with
\massdivp\ and \ordzero\ we learn that (up to
a phase):
\eqn\eqprds{
\Psi_{grav}^{HET}(y^a) = \Psi_{grav}^{II}(t_\alpha)
}
on very general grounds.

\newsec{Gravitational correction for the FHSV model}

Computation of the
integral expression \gc\ leads to
interesting automorphic forms
associated to threshold corrections.
In this section we work in reverse:
we use a known result on automorphic
forms to determine a threshold correction.

We will consider the FHSV model
introduced in \fhsv. A discussion of phase transitions in this model and a
careful discussion of some topological subtleties in its definition
can be found in \aspintran. The model consists of  an $N=2,d=4$
string dual pair where the type IIA theory is
formulated on a Calabi-Yau manifold of
the form $X_3=(T^2 \times S)/\IZ_2$ where $S$ is a
$K3$ surface which   double
covers  an Enriques surface. The
$\IZ_2$ acts as $z \rightarrow -z$ on $T^2$ and as
the free  Enriques involution on $S$. In addition there
are some discrete degrees of freedom which
must be included \refs{\fhsv,\aspintran}. The resulting Calabi-Yau
space has Betti numbers $h^{1,1}(X_3)=h^{2,1}(X_3)=11$ and hence
$\chi(X_3)=0$.

An important property of this model is that
the exact vectormultiplet moduli space is  a quotient of:
\eqn\localvm{
SL(2,\IR)/SO(2) \times  O(10,2)/[O(10)\times O(2)]
}
with its natural K\"ahler metric.
That is, the special K\"ahler geometry of the
moduli space \localvm\ does not receive string quantum corrections
because of the decoupling of vector and hypermultiplets and it was
argued in \fhsv\ that it is also uncorrected by string world-sheet
quantum corrections from world-sheet instantons.  It thus follows
that $\CF \sim S y^2$ and
$K=- \log[Re S] - \log[-(Im y)^2] $ exactly.
Nevertheless, there will be nontrivial quantum corrections
for other $F$-terms in the low energy
effective theory. The relevant automorphic
form for the $R^2$ coupling is described in
the next section.

The global identifications are associated with
duality transformations, and can be determined
from the lattice of RR charges. It was shown in
\refs{\fhsv,\aspintran} that the moduli space is:
\eqn\fhsvvm{
\CM_{\rm vm}^{\rm quantum} = SL(2,\IZ) \backslash SL(2,\IR)/SO(2) \times O(M)
\backslash O(10,2)/[O(10)\times O(2)] .
}
Here $O(M)$ is the group of automorphisms
of
\eqn\gamdef{
M \equiv  E_8(-2) \oplus II^{1,1}(2) \oplus II^{1,1}(1)
}
(this is the lattice denoted $M$ in \borchiv). Here $II^{1,1}$
is the even self-dual Lorentzian lattice and for any self-dual
lattice $L$, $L(n)$ denotes the lattice with norm-squared
 rescaled by $n$. \foot{The sign convention in \gamdef\ is
chosen to agree with the standard convention of algebraic geometry}
 $M$ may be interpreted in many ways.
One way is that it is the sublattice of integral cohomology
$H^*(S;\IZ)$ even under the Enriques involution.

\subsec{Determining $F_1$}

In order to determine $F_1$ we first use the
representation \bcoviii\ from the type II side.
Note first that the dependence on the Kahler
moduli of $T^2$ can only come from the
untwisted unprojected sector of the $\IZ_2$ orbifold.  The calculation in
the untwisted sector is,
up to a factor of $1/2$,
 the same    as the
calculation on a Calabi-Yau manifold
of the form
$T^2 \times S$ where $S$ is a
K3 surface. For the latter manifold one easily
finds\foot{The physical implications of this are explored in
\hmiii.}:
\eqn\teekay{
F_1^{\rm kahler} = - 24 \log \parallel \eta^2(T) \parallel^2
}
where $T$ is the Kahler modulus of $T^2$,
the factor of
$24$ comes from the elliptic genus
of K3, and the T-duality invariant norm is
$ \parallel \eta^2(T) \parallel^2 \equiv
Im T \vert  \eta^2(T)  \vert^2$. To compare this to the discussion
in sec. 3.1 we view this as a $K3$ fibration where the $\IP^1$ base
is $T^2/\IZ_2$. Since the area of this $\IP^1$ is half that of $T^2$
we should identify $t_s = T/2$.
Taking into account this identification and the factor of two from
the orbifold we have:
\eqn\fhsvgc{
F_1 =  \log \parallel { 1 \over  \eta^{24}(2t_s) \Psi^{II}}
\parallel^2
}
where $\Psi^{II}$ is a section of a line bundle
over
$O(M)
\backslash O(10,2)/[O(10)\times O(2)]$, or,
equivalently, an automorphic form
for $O(M)$. We will now use our
knowledge of the physics to determine
the weight of the form and its singularities.

The weight is easily determined from
\weight\ to be $w=4 - \chi /12 = 4$.
The singularities of $\Psi^{II}$ must come from
singular Calabi-Yau manifolds $X_3$. As explained in
\aspintran\  $X_3$ becomes singular when

\item{1.}  $S$ develops an ADE singularity, or

\item{2.} $S$ develops a quantum singularity, which by mirror symmetry
can be viewed  as a singularity occuring at points in the
moduli space where the $\IZ_2$ fails to act freely.
As explained in \aspintran, the
singularities of the second kind occur at
the divisor in $O(10,2)/O(10) \times O(2)$
orthogonal to a norm $-2$ vector.
\foot{This is the divisor denoted
$\cup H_d$ in \borchiv. As shown in that
paper, there is only one $O(M)$
orbit.}

Physically, the singularities of the first
kind correspond to theories with enhanced
$N=4$ supersymmetry. At these points we have
$\Delta n_H - \Delta n_V = 0$. There is thus
no singularity in $\Psi$ at such points.
On the other hand, singularities of
the second kind correspond to enhanced
$N=2$, $SU(2)$, $N_f=4$ theories.
Using \zeros\ we see that along
 this divisor $\Psi$ will have a zero of
order
\eqn\finzero{
{1 \over  6} [ 8 - 2] = 1 .
}

The above properties of $\Psi$ are also
easily reproduced within the dual heterotic
theory. The heterotic dual is obtained as
an asymmetric $\IZ_2$ orbifold for lattices in
$\IR^{22,6}$ with orthogonal decomposition:
\eqn\hetlatt{
\Gamma^{9,1} \oplus \Gamma^{9,1} \oplus \Gamma^{1,1}
\oplus \Gamma^{1,1} \oplus \Gamma^{2,2}
}
where the $\IZ_2$ acts as
$ \vert P_1, P_2, P_3, P_4, P_5 \rangle
\rightarrow e^{ 2\pi i \delta \cdot P_3}
\vert P_2, P_1, P_3, -P_4, -P_5 \rangle$ with $\delta$ the
order two shift vector defined in \fhsv.
The weight is now determined from \autweight\
and the enhanced symmetry divisors
correspond to the $N=4$ and $N=2$
theories described above \fhsv.
Since the vector multiplet moduli space is uncorrected
there is no possible singularity in
$\tilde S = S$, and we recover the above
result on the zeros of $\Psi_{\rm grav}^{HET}$.

In principle the correction $F_1$ could be obtained
by evaluating \gc\  explicitly as in
 \refs{\hm, \tkawai, \henning, \lustline}.
However, this is not necessary since
enough is known about the relevant modular
forms to determine $F_1$ without calculation.
As we have mentioned,
by the Koecher principle, a nonsingular
modular form for $O(M)$ is determined
by its weight and the order of its zeros
on $O(10,2)/[O(10) \times O(2) ] $. The reason is
that two forms of the same weight and zeroes
would have a ratio which is an
$O(M)$-invariant holomorphic function
$\Psi_{12} = \Psi_1/\Psi_2$ on
$O(10,2)/[O(10) \times O(2) ] $. This would
descend to the quotient space
$O(M) \backslash O(10,2)/[O(10) \times O(2) ] $.
However, this space has a compactification
(the ``Baily-Borel compactification'') by
adding varieties of  {\it dimension} zero and one
\ref\sterk{H. Sterk, ``Compactifications of the
period space of Enriques surfaces. Part I,''
Math. Z. {\bf 207}(1991)1}
\ref\scattone{F. Scattone,{\it On the compactification
of moduli spaces for algebraic K3 surfaces},
Mem. of the AMS, {\bf 374} (1987)}.
Hence,
by Hartog's theorem $\Psi_{12}$ must be a constant.

It turns out that an automorphic form of precisely
the
required weight and with precisely the required
zeros has already been constructed by
Borcherds in \borchiv. We call this form
$\Phi_{BE}$, the
Borcherds-Enriques form. It may be   defined
as follows.

We first make a choice of null vector
 $v\in M\cong
[\Gamma^{9,1} \oplus \Gamma^{9,1}]^{\IZ_2}\oplus II^{1,1}(1) $.
\foot{This choice of null vector corresponds to
a choice of which integers we use for Poisson
resummation in the calculation of \hm.} Here the $\IZ_2$ superscript
indicates that we take the   part of the lattice  invariant
under the exchange of the two factors.
We next define  $L = (v^\perp/v\IZ)$.
In particular,  we make the choice of a primitive
null vector in $II^{1,1}(2)$ so that
$L\cong    E_8(-2)\oplus \Gamma^{1,1}(1) $.  Let
$r>0$ denote lattice vectors in the forward light-cone.
Denoting vectors in the lattice by
$r=(\vec b; m,n) \in L$ and  points
in a tube domain by $y\in \IR^{1,9} + i C^+$
we define \borchiv:
\eqn\boren{
\eqalign{
\Phi_{BE}(y) &\equiv
e^{2\pi i \rho\cdot y}
\prod_{r>0}(1-e^{2\pi i r\cdot y})^{(-1)^{m+n} d(-r^2/2)}
\cr
\sum d(n) q^n & = q^{-1} {\prod (1+q^{2n+1})^8\over
\prod (1-q^{2n })^8} \cr}
}
where $\rho=(0;0,1)$ is a Weyl vector.
It is shown in \borchiv\ that the
form $\Phi_{BE}(y)$ satisfies
two key properties:

\item{1.}  $\Phi_{BE}(y)$ is an automorphic form
on $O(10,2)/[O(10)\times O(2)]$ for the
discrete group $O(M)$ of weight 4.

\item{2.}  The zeroes  of $\Phi_{BE}(y)$ are first
order and are located on the divisor of
$O(10,2)/O(10) \times O(2)$ of negative definite
subspaces orthogonal to norm $-2$ vectors in
$M$.

As we have remarked several times, these
two properties determine $\Phi$ up to a constant,
so we can finally obtain the full nonperturbative
result for the gravitational threshold correction:
\eqn\fhsvfnl{
F_1 =  \log \parallel { 1 \over  \eta^{24}(2 \tau_S) \Phi_{BE}(y) }
\parallel^2
}
where we have written the final answer in terms of the heterotic
dilaton $\tau_S$.

The factor of $2 \tau_S$ in \fhsvfnl\ has physical significance
and is quite closely related to
a similar factor in the discussion of $N=2$ gauge theory with
matter given in \swb. Expanding \fhsvfnl\ gives a term linear
in $S$ and a power series in $q_S^2$. The latter should be
interpreted as coming from zero size instantons as in \hmiii.
Since a single instanton has action $q_S$ this indicates that
only even numbers of instantons contribute to $F_1$. A similar
result was found in \swb\ for $N=2$ theories with matter as
a result of an anomalous $\IZ_2$ symmetry identified with the
center of the global $O(2 N_f)$ flavor group. In that situation
amplitudes with odd numbers of instantons were odd under the $\IZ_2$
and thus did not contribute to the even metric on moduli space.
In our situation there is also a natural $\IZ_2$, it is the $\IZ_2$ which
is $+1$ on all untwisted states of the $\IZ_2$ orbifold and $-1$
on all twisted states. The above result tells us that again
odd numbers of
instantons are odd under this $\IZ_2$ and hence do not contribute to
$F_1$ which is even. Note also that at enhanced
symmetry points with $N=2$ supersymmetry and gauge group $SU(2)$
with $N_f=4$ the four matter multiplets arise from the twisted
sector of the orbifold and thus are odd under the $\IZ_2$.
The orbifold $\IZ_2$ thus
agrees with the $\IZ_2$ of \swb\ when restricted to the $N=2$
gauge sector.

\newsec{Application 1: K3 determinants}

The dependence of $F_1$ on {\it complex structures}
was determined in \bcov\ to be given by a
combination of Ray-Singer torsions.
Specifically we have:
\eqn\rst{
F_1^{\rm complex} = + \sum_{0\leq p,q\leq 3} p q (-1)^{p+q} \log \Det'
\Delta_{\pb}^{p,q}
}
We have determined this quantity for
$X_3 = (T^2 \times S)/\IZ_2$. Using
special properties of this manifold we
can extract a relation between
$\Phi_{BE}$ and determinants of $\pb$
operators on K3 surfaces. Since
$X_3$ is a self-mirror family we can
work equally well with Kahler or complex
moduli, a fact which will be exploited in
the next section.

Using the Hodge dual  we obtain an
isospectral isomorphism
$\Omega^{p,q} \cong \Omega^{q,p}$
and hence:
$\Det' \Delta_{\pb}^{p,q} =
\Det' \Delta_{\pb}^{q,p}$. Moreover,
using the covariantly constant
$3$-form we have a second
isospectral isomorphism:
$\Omega^{0,q} \cong \Omega^{3,q}$.
Finally, let $\Omega^{1,q}_{\perp}$
denote the space of forms orthogonal
to the  harmonic forms. We can obtain
the isospectral isomorphism:
\eqn\ispctrl{
\eqalign{
\Omega^{1,q}_{\perp} \quad \cong \quad &
  \pb(\Omega^{0,q}) \oplus \pb^\dagger (\Omega^{2,q}) \cr
 \quad \cong \quad &
\pb^\dagger (\Omega^{3,q}) \oplus
\pb(\Omega^{1,q} )
 \quad \cong \quad  \Omega^{2,q}_{\perp} \cr}
}
and hence
 $\Det' \Delta_{\pb}^{p,q} =
\Det' \Delta_{\pb}^{3-p,q}$.
It follows that:
\eqn\fiii{
F_1 = 9  \lgdt{0}{0} -6  \lgdt{1}{0}
+ \lgdt{1}{1}.
}

Moreover, the metric is a product on
$T^2 \times S$ where $T^2$ is the
torus and $S$ is the K3
surface   which double-covers the
Enriques surface. Thus,
$\Delta = \Delta_{T^2} + \Delta_{S}$
and the Laplacian is block diagonal if
we decompose the space of $(p,q)$
forms according to:
\eqn\decom{
\Omega^{p,q}((T^2 \times S) )
\cong \oplus_{p',q'}
\Omega^{p-p',q-q'} (T^2) \otimes  \Omega^{ p', q'} (S)
}
Also, the involution $(-1, I_{Enriques})$ on
$T^2 \times S$ commutes with
the Laplacian so we can decompose all
spaces into the $\pm$ eigenspaces under
the involution.

We also need some facts about the
spectrum of the Laplacians $\Delta^{p,q}_{S}$ and
their determinants which we define using zeta-function regularization.
The spectrum of the Laplacian on 0-forms
leads to a zeta-function which can naturally
be written as a sum of zeta functions for the
even and odd functions on $S$:
\eqn\zetakiii{
\zeta_{S}(s) = \zeta_{S}^+(s) + \zeta_{S}^-(s)
}
All the other relevant $\zeta$ functions for
$S$ can be written in terms of $\zeta^\pm_S$.
Using the fact that the covariantly constant
$(2,0)$ form on $S$ is odd one can derive
the relevant zeta functions for the Laplacian
acting on other forms. In particular:
\eqn\othzets{
\eqalign{
\Omega^{1,0}_{\pm}: ~
& \quad  \zeta_{S}^+(s) + \zeta_{S}^-(s) \cr
\Omega^{1,1}_{\pm}: ~
& \quad  2\zeta_{S}^+(s) + 2\zeta_{S}^-(s) \cr}
}

We now consider the $\zeta$-function on
$\Omega^{p-p',q-q'}_\pm (S)
\otimes  \Omega^{ p', q'}_\pm (T^2) $.
The spectrum of the Laplacian is the sum of
pairs $\lambda_i^\pm(S) + \lambda_j^\pm(T^2)$.
In fact,
we know the spectrum $\lambda_j(T^2)$ of the flat torus
quite explicitly, so the $\zeta$-function is of the
form
\eqn\mixd{
\sum_{i,n_1, n_2} \Biggl(\lambda_i^\pm (S) + {\vert n_1 \tau + n_2\vert^2
\over \tau_2^2} \Biggr)^{-s}
}
It is quite important to separate out the
terms corresponding to
 a zeromode on  $T^2$ or $S$ from expressions
such as \mixd.  Let us define the functions
\eqn\nonzms{
\eqalign{
\mu^+(s) &  = \sum_{\lambda_i^+\not=0 ,(n_1, n_2)\not=0} \Biggl(\lambda_i^+
(S) + {\vert n_1 \tau + n_2\vert^2 \over \tau_2^2} \Biggr)^{-s} \cr
\mu^-(s) &  = \sum_{\lambda_i^-\not=0 ,(n_1, n_2)\not=0} \Biggl(\lambda_i^+
(S) + {\vert n_1 \tau + n_2\vert^2 \over \tau_2^2} \Biggr)^{-s} \cr}
}
Then, a short calculation produces the
following $\zeta$-functions
\eqn\morezets{
\eqalign{
\Omega^{0,0}_+(X_3) & \qquad \zeta^+_{T^2}(s) + \zeta^+_{S}(s) + \mu^+ + \mu^-
\cr
\Omega^{1,0}_+(X_3) & \qquad \zeta^-_{T^2}(s)
+ \zeta^+_{S}(s) + 2 \zeta^-_{S}(s) + 3\mu^+ + 3\mu^- \cr
\Omega^{1,1}_+(X_3) & \qquad (n_e + n_0 +1)\zeta^-_{T^2}(s)
+ 5\zeta^+_{S}(s) + 4 \zeta^-_{S}(s) + 9\mu^+ + 9\mu^- \cr}
}
where $n_e, n_0$ are the dimensions of the
even and odd harmonic $(1,1)$-forms. For our
example these are $n_e=n_0=10$.
Computing the combination
in \fiii\ we find that the mixed eigenvalues
in $\mu^\pm$ cancel out and we are left with
\eqn\finalzet{
24 \zeta^+_{T^2} + 8 (\zeta^+_{S} -  \zeta^-_{S})
}
and we recover the fact that $F_1$ can be written
as a product of the function for $T^2$ and a function
for $S$. Moreover, we obtain the result
\eqn\diffdet{
8 [\log \Det' \Delta^{0,0}_+ - \log \Det' \Delta^{0,0}_-] = - \log \parallel
\Phi_{BE}\parallel^2
}
This result is closely related to a result of Jorgenson and Todorov \jtii.

\newsec{Application 2: Mirror maps and curve counting}

In $F$-compactification to 8 dimensions
the Narain moduli space
$\CN^{2,18}$ plays an important role
\effcomp.
On the heterotic side it is natural to describe
this space in terms of tube domain coordinates
as $t^i \in \IR^{1,17} + i C^+$. On the F-theory
side one parametrizes the space in terms of
the positions $z_i\in \IP^1$ where $7$-branes
have been inserted. Equivalently the
elliptic fiber of the elliptically fibered $K3$ degenerates
at points $z_i$. It would be very interesting to
determine the exact mapping between the
$z_i$ and the $t^i$. If we restrict to the
$O(10,2)$ subspace corresponding to
Enriques double-covers (all of which are
elliptically fibered) then this is simply the
mirror map. The above results give some
nontrivial information on this map.
(See
\ref\lustmap{
G. L. Cardoso,  G. Curio,  Dieter Lust,  T. Mohaupt,
``On the Duality between the Heterotic String and F-Theory in 8 Dimensions,''
hep-th/9609111} \
for an explicit determination of the map on a
two-parameter subspace.)

The proper interpretation of \bcovf\  is that
$\Psi_{\rm grav}^{II}(t)$ is a section of
bundle with a hermitian metric.
Using special coordinates and the
mirror map we can define a trivialization
near a large K\"ahler structure boundary
point in terms of the fundamental
holomorphic period
$\varpi_0$. Then if $t(z)$ is the mirror map:
\eqn\borlaugh{
( \varpi_0)^{3 + h_{1,1} - \chi/12}
\det (\p t/\p z) \Psi_{\rm grav}(t)
}
is an invariant object: it is the quotient of two sections.
On the other hand, we know that its only
singularities can occur on the discriminant locus.
Hence, in the
case of the Enriques surfaces,
 we expect \borlaugh\  with
$ \Psi_{\rm grav}(t)= \Phi_{BE}(t) $ to be
proportional to $\prod_{ij}(z_i-z_j)^{n_{ij}} $
for some integers $n_{ij}$. The
integers $n_{ij}$ can be determined by
counting elliptic curves near the
discriminant locus. Indeed, once
$t(z)$ is known one could then proceed to
count elliptic curves in the Calabi-Yau
$X_3$ (and hence extract elliptic curves in
the Enriques surface) using the result \bcov\bcova:
\eqn\ellcount{
\eqalign{
- \log\bigl[
( \varpi_0)^{3 + h_{1,1} - \chi/12}
\det (\p t/\p z) \Psi_{\rm grav}(t)
\bigr]  & = - { 4 \pi i  \over  12} \int_{X_3} c_2\wedge  \omega \cr
- 2 \sum_{\Sigma \in H_2(X_3;\IZ) } n^{(1)}(\Sigma)
\log\bigl[ \eta(e^{2 \pi i \int_\Sigma \omega} )\bigr]
&
- { 1 \over  6} \sum_{\Sigma \in H_2(X_3;\IZ)} n^{(0)}(\Sigma)
\log\bigl[ 1- e^{2 \pi i  \int_\Sigma \omega} \bigr] \cr}
}
where $n^{(1)}, n^{(0)}$ are the number of
elliptic and rational instantons and $\omega$ is
the complexified K\"ahler class. Since these are not
isolated the counting of these instantons is
subject to the standard caveats.

\newsec{Conclusions and Conjectures}

The main motivation for this paper was
the hope
that the present result will lead to a
better understanding of the relation between
the BPS algebras discussed in  \hmalg\ and
threshold corrections.
Given the result \fhsvgc\
we are led to the following conjecture.

{\it  Conjecture}: For heterotic compactification
on $K3\times T^2$ we
always have
\eqn\conjec{
F_1 = \log \parallel  {1 \over q_{\tilde S}^2 \Phi(y)}
(1 + \CO(q_S)) \parallel^2
}
where $\Phi(y)$ is the denominator product for
a GKM algebra closely related to the algebra of perturbative
BPS states.

Understanding the precise relation between the algebra
associated with the denominator product and the
algebra of BPS states remains elusive.
In the FHSV model considered here $\Phi(y)$ is the denominator
product for the ``fake monster Lie superalgebra'' \borchmoon. (It can
also be interpreted as the trace of a involution in the
Monster in the Lie algebra cohomology of the Monster
Lie algebra.)  The algebra of BPS states in the FHSV
model resembles a generalized Kac-Moody algebra,
but is graded by $\Gamma^{10,2}$. Our result
strongly suggests that some subquotient of the algebra of BPS states, perhaps
associated to cohomology associated with
a null direction, will be the
``fake monster super Lie algebra.''
It would be very interesting to make this
suggestion concrete, but we have not done so.
In this connection it is intriguing to note that by
expanding the same form around different cusps
one can obtain {\it distinct} GKM's \borchvii.

On a more speculative note, it is tempting to conjecture that
 \conjec\ is the first term
in an expansion in powers of $q_S$ of an automorphic form which
governs the exact, non-perturbative gravitational threshold correction
in all $N=2$ theories: $F_1 =- \log \parallel \Phi \parallel^2$, and that
$\Phi$  is associated to
an electric or magnetic subalgebra of the full
algebra of BPS states associated to the
Calabi-Yau space \hmalg.

Finally, an interesting direction for future research
is to understand if there is a connection between
  BPS algebras and higher dimensional
current algebras. There is a hint of such a connection
since the quantity $F_1$ also appears in
\clash.

\bigskip

\centerline{\bf Acknowledgements}\nobreak

We would  like to thank   R. Borcherds,
M. Douglas, A. Gerasimov, J. Jorgenson,
A. Losev, D. Morrison, N. Nekrasov,
S. Shatashvili, A. Strominger,  A. Todorov,
and G. Zuckerman for discussions.
We are extremely grateful to J. Louis for extensive
correspondence on gravitational couplings and
holomorphic anomalies, and for comments on
the manuscript. We are also grateful to  V. Kaplunovsky
for an important discussion on this topic.

GM would like to
thank  the Aspen Center for Physics for providing
a stimulating atmosphere during the beginning
of this work. This work was  supported in part
by NSF Grant No.~PHY 91-23780 and
DOE grant DE-FG02-92ER40704.

\listrefs
\bye